\documentstyle[preprint,aps]{revtex}
\begin{document}
\draft
\preprint{UM-P-93/81, OZ-93/21}
\title{Effective Lagrangian Approach to the Fermion Mass Problem}
\author{D. S. Shaw and R. R. Volkas}
\address{Research Centre for High Energy Physics, School of Physics,\\
University of Melbourne, Parkville 3052, Australia.}
\maketitle

\begin{abstract}

An effective theory is proposed, combining the standard gauge
group $SU(3)_{C}\otimes SU(2)_{L}\otimes U(1)_{Y}$ with a
horizontal discrete symmetry. By assigning appropriate charges under
this discrete symmetry to the various fermion fields and to (at least)
two Higgs doublets, the broad spread of the fermion mass and mixing
angle spectrum can be explained as a result of suppressed,
non-renormalisable terms. A particular model is constructed which
achieves the above while simultaneously suppressing neutral
Higgs-induced flavour-changing processes.

\end{abstract}

\section{Introduction and Philosophy}
\label{intro}

One of the most intriguing problems still outstanding with the
standard model (SM) is the unexplained nature of the quark and lepton
mass and mixing
angle hierarchies. The range of values for the mixing angles
spans some three orders of magnitude, while that for the masses
spans at least six, yet in the SM these values result from Yukawa
terms of the same form but with different (and unpredicted) coupling
constants. It seems very unnatural for the range of values for these
constants to span so many orders of magnitude. One way to explain
this hierarchy is to propose that, through some as yet unknown
mechanism, the heavier particle masses and larger mixing angles are
generated at lowest order in some expansion parameter, and that the
particles of smaller mass receive no lowest-order contribution, only
gaining masses from higher-order terms.

In this paper we outline a mechanism by which this cascade effect
may be achieved. The key points are:

(1) We assume a more fundamental theory to exist
at very high energies (we expect this level to be in the TeV range)
which contains much more information on quark and lepton flavours than
does the SM.

(2) We further assume that the effective theory below the electroweak
scale ($\simeq 300$ GeV) contains the SM gauge symmetry
$G_{SM} = SU(3)_{C}\otimes SU(2)_{L}\otimes U(1)_{Y}$
{\it together with an additional horizontal discrete symmetry $D$\ }
[it is, of course, possible to try and explain the mass and mixing
angle spectrum just using a horizontal symmetry, without recourse to
effective theory. For examples of this, see (using discrete symmetries)
\cite{discref} and (using continuous symmetries) \cite{contref}].
This horizontal discrete symmetry is the minimal amount of information
on the flavour sector of the theory that is assumed to trickle down
from the fundamental theory to our effective low-energy world. Beyond
this, no further information about the high-energy theory is required
nor sought.

(3) The effect of the discrete symmetry $D$ is to put non-trivial
flavour structure into the pattern of higher-dimensional,
non-renormalisable operators that will contribute to fermion mass
generation after electroweak symmetry breaking. The effective theory
will contain terms of the form
\begin{equation}
\label{term-types}
\overline{f_{1}}f_{1}\phi,\;\;
\frac{1}{\Lambda^{2}}\overline{f_{2}}f_{2}\phi(\phi^{\dag}\phi),\;\;
\frac{1}{\Lambda^{4}}\overline{f_{3}}f_{3}\phi(\phi^{\dag}\phi)^{2}.
\end{equation}
where $\Lambda$ sets the scale of the new flavour physics.
When $\phi$ gains a non-zero VEV
$\langle\phi\rangle \equiv v$ the above fermions gain a
hierarchical pattern of masses,
\begin{equation}
m_{1} \sim v \gg m_{2} \sim \frac{v^{3}}{\Lambda^{2}} \gg m_{3} \sim
\frac{v^{5}}{\Lambda^{4}} \gg \ldots
\end{equation}
because we assume that $\Lambda \gg v$. Hierarchical mixing terms
between flavours will be arranged to also reproduce a CKM mixing angle
hierarchy (a related but contrasting approach postulates the radiative
generation of mass and mixing angle hierarchies. See \cite{Babu&Ma}
for a review).

(4) At least two Higgs doublets are required in order for operators
like $\phi^{\dag}\phi$ to transform non-trivially under $D$.

Our analysis will uncover simple candidates for the horizontal
discrete symmetry $D$ that yield a reasonable qualitative
understanding of masses and mixing angles\footnote{While this paper
was being prepared, a paper on a similar
theme appeared by Leurer, Nir and Seiberg \cite{singlet}. The above
paper also brought our attention to an earlier paper by Froggatt and
Nielsen \cite{singref} which advocates a very similar idea.}.

Before proceeding with our analysis, we would like to make a few more
points: (i) The new flavour physics is assumed to be encoded by a new
discrete symmetry, rather than a gauge or continuous global symmetry.
This may be phenomenologically advantageous, because the new symmetry
has to remain unbroken until the electroweak scale. This would, in
general, not be tolerable for a horizontal gauge symmetry, and the
breaking of a continuous global symmetry would, in general, produce
troublesome Goldstone Bosons\footnote{In particular models employing
continuous symmetries it may, however, turn out that dangerous
processes are sufficiently suppressed by small mixing angles.}.
(ii) The dimensionless coefficients of
the operators in Eq.\ (\ref{term-types}) are assumed to be numbers of
order 1. Exactly what constitutes a number of order 1 is a subjective
matter. Our opinion is that any number from about 0.2 to 5 qualifies
as such. (iii) Note that we assume the flavour hierarchy structure to
be due to two things: the new symmetry $D$ and the additional Higgs
doublets. An effective lagrangian can have structure without these
two things, but then all of this structure would be due to the unknown
high-energy theory. For instance, some fundamental theory may well
yield hierarchical values for the coefficients of dimension-4
$\overline{f}f\phi$ terms, thus explaining flavour. However such
theories are inaccessible to us at present, so we concentrate on the
alternative and more interesting possibility that the flavour
information is already present at the electroweak scale. (iv) We
expect flavour-changing neutral Higgs effects to exist in the
effective theory. This phenomenological signature is an
important generic prediction of models of our form.

The remainder of this paper is structured as follows: In
Sec.~\ref{gen-gap} we examine the possibility that the third
generation fermions ($t$,$b$,$\tau$) gain mass from dimension-4
operators, while second generation ($c$,$s$,$\mu$) and first
generation ($u$,$d$,$e$) fermions gain mass from dimension-6 and
dimension-8 operators respectively. Sec.~\ref{full-hier} then
considers a more complicated pattern wherein only the top quark gains
mass at the dimension-4 level. We make some phenomenological remarks
on Higgs boson physics in Sec.~\ref{fcncs} and we conclude in
Sec.~\ref{conc}.

\section{The Generation Gap}
\label{gen-gap}

In this section we give a simple, warm-up example of how our mechanism
is implemented and use it to propose an explanation of the most
prominent trend in the mass spectrum. A quick look at the fermion
masses will reveal a definite hierarchy
between the three generations of particles. In each particle sector
(the ``up'' quarks, the ``down'' quarks and the charged
leptons), the third generation particle is consistantly heavier than
the second generation particle which in turn is heavier than the first
generation particle (we reserve the term ``generations'' here to refer
to the repeated patterns of fermion quantum numbers, so that the up
and down quarks and the electron and electron neutrino constitute the
first generation, the charm and strange quarks and the muon and its
neutrino the second generation and so on). This separation in scale is
typically around two orders of magnitude per generation. For
simplicity, it is assumed that
the three sectors of particles will follow identical patterns of mass
generation, so that only one ``generic'' sector need be looked at to
know how all the fermions will behave (for ease of reference, this
sector will be named for the charged leptons: tauon, muon and
electron).

The mass Lagrangian is assumed to be made up of the following types of
terms:

\begin{equation}
\label{mass-Lag}
{\cal L}_{\rm mass} = \lambda\overline{f}_{L}f_{R}\phi_{1} +
\frac{\lambda}{\Lambda^{2}}\overline{f}_{L}f_{R}\phi_{1}
(\phi_{2}^{\dag}\phi_{1}) + \frac{\lambda}{\Lambda^{4}}
\overline{f}_{L}f_{R}\phi_{1}(\phi_{2}^{\dag}\phi_{1})
(\phi_{2}^{\dag}\phi_{1}) + {\rm H.c.}
\end{equation}
The first term is a standard, renormalisable Yukawa coupling term, and
it is assumed that the heaviest particle(s) (here the tauon) will gain
mass(es) from terms such as this. The remaining terms are
non-renormalisable terms. The constant $\Lambda$ (which has the
dimensions of mass) is required to ensure that the Lagrangian density
has mass dimension 4. The value of $\Lambda$ is of the order of the
breaking energy for the
(unknown) high-energy fundamental group which is broken down to a
product of the SM and some discrete group $D$, and thus is very
large (of the order of TeVs). This will suppress these higher-order
(in $\Lambda$) terms. The muon is
presumed not to participate in any of the renormalisable (tree-level)
terms, but to get its mass only from the non-renormalisable terms of
order $\Lambda^{2}$ or above. The electron only combines in terms
such as the third term in Eq.\ (\ref{mass-Lag}) above (and in
higher-order terms),
thereby receiving only a very small mass [suppressed heavily by the
factor $(\frac{v}{\Lambda})^4$].

In order to assure that this can happen, the charges of the fermion
fields and the Higgs fields under the discrete symmetry $D$ are chosen
so that no undesired terms (such as a tree-level, dimension-four
mass term for the electron) can be found which are invariant under $D$
while it is assumed that the effective Lagrangian is required to be
invariant under the discrete symmetry.

For the current example, the simplest
discrete group that will provide a mass hierarchy between the
generations is $Z_5$. It is easy enough then to restrict the particle
masses to appropriate sizes by giving them suitable charges under
$Z_5$, however careful choices need to be made if a reasonable form
for the Cabbibo-Kobayashi-Maskawa (CKM) matrix is to be obtained.
A passable approximation to the CKM matrix can indeed be achieved
using $Z_5$, but at the cost of introducing a fine-tuning condition.
The CKM matrix is comprised of rotation matrices, one from each of the
two quark sectors, used to diagonalise the quark mass matrices. The
fine tuning condition arises in the process of determining these
rotation matrices, where a very precise cancellation is required
between two (a priori) large, distinct parameters in order to
determine a consistant form for the rotation matrices.

Since the simplest possibility for the discrete group runs into
trouble, a more complicated choice is sought. It turns out that the
simplest choice of discrete group that can achieve the same results as
for the $Z_{5}$ symmetry discussed above, but with less egregious
fine tuning, is the next smallest cyclic group, $Z_{6}$.
Table \ref{zed6} lists the various particle fields and the charges
assigned to them under the discrete symmetry.

The only 4-dimensional term allowed by the above assignments is

\begin{equation}
{\cal L}_{\rm 4-d} = \lambda_{1}\overline{\tau}_{L}\tau_{R}\phi_{1} +
{\rm H.c.}
\end{equation}
generating a mass for the tauon. The possible 6-dimensional terms are
\begin{eqnarray}
{\cal L}_{\rm 6-d} & = & \frac{\lambda_{2}}{\Lambda^{2}}
\overline{\tau}_{L}e_{R}\phi_{1}(\phi_{2}^{\dag}\phi_{1}) +
\frac{\lambda_{3}}{\Lambda^{2}}
\overline{\mu}_{L}\mu_{R}\phi_{2}(\phi_{1}^{\dag}\phi_{2}) +
\frac{\lambda_{4}}{\Lambda^{2}}
\overline{\tau}_{L}\mu_{R}\phi_{1}(\phi_{2}^{\dag}\phi_{1})
+ \nonumber \\
 & & \frac{\lambda_{5}}{\Lambda^{2}}
\overline{\mu}_{L}e_{R}\phi_{2}(\phi_{1}^{\dag}\phi_{2}) + {\rm H.c.}
\end{eqnarray}
These terms will add small contributions to the tauon mass,
generate the muon mass and some of the mixing angles. The
electron mass and the rest of the mixing angles will be
generated from the following 8-dimensional terms:

\begin{eqnarray}
{\cal L}_{\rm 8-d} & = &
\frac{\lambda_{6}}{\Lambda^{4}}\overline{\mu}_{L}\tau_{R}\phi_{2}
(\phi_{1}^{\dag}\phi_{2})^{2} +
\frac{\lambda_{7}}{\Lambda^{4}}\overline{e}_{L}e_{R}\phi_{2}
(\phi_{1}^{\dag}\phi_{2})^{2} +
\frac{\lambda_{8}}{\Lambda^{4}}
\overline{e}_{L}\mu_{R}\phi_{2}(\phi_{1}^{\dag}\phi_{2})^{2}
+ \nonumber \\
 & & \frac{\lambda_{9}}{\Lambda^{4}}
\overline{e}_{L}\tau_{R}\phi_{1}(\phi_{2}^{\dag}\phi_{1})^{2} +
{\rm H.c.}
\end{eqnarray}
There will also be very small corrections from higher order terms
(dimension 10 or above).

Together, these terms will result in a mass matrix of the form
\begin{equation} M = \left( \begin{array}{ccc}
\mu & \mu & \mu \\
m & m & \mu \\
m & m & M
\end{array} \right) , \end{equation}
where $M \sim \lambda v \gg m \sim \lambda v (\frac{v}{\Lambda})^2 \gg
\mu \sim \lambda v (\frac{v}{\Lambda})^4$ (note that these values are
orders of magnitudes only, so that different instances of the same
value may differ by small correction factors which will result from
higher-order contributions and from differing Yukawa coupling
constants). In order to find the mixing
angles for this model, we need to find the rotation matrices that will
diagonalise the above matrix. In general, two such matrices, $R$ and
$L$, are required ($D$ is the diagonalised mass matrix):

\begin{equation}
L^{\dag}MR = D.
\end{equation}
However, only the left-hand rotation matrix, $L$, is needed to
determine the mixing angles. If we multiply the raw mass matrix above
by its hermitian conjugate, the resulting ``squared'' matrix will be
diagonalised by L alone (which we can thus determine by studying the
characteristic equation for the ``squared'' matrix):

\begin{equation}
D^{2} = L^{\dag}MM^{\dag}L.
\end{equation}
This ``squared'' matrix has the form

\begin{equation}
MM^{\dag} = \left( \begin{array}{ccc}
3\mu^{2} & \mu^{2} + 2m\mu & 2m\mu + M\mu \\
\mu^{2} + 2m\mu & 2m^{2} + \mu^{2} & 2m^{2} + M\mu \\
2m\mu + M\mu & 2m^{2} + M\mu & M^{2} + 2m^{2}
\end{array} \right) \sim \left( \begin{array}{ccc}
\mu^{2} & m\mu & m^{2} \\
m\mu & m^{2} & m^{2} \\
m^{2} & m^{2} & M^{2}
\end{array} \right)
\end{equation}
(noting that, in terms of order of magnitude, $M\mu \sim m^{2}$)
and diagonalising this matrix leads to the (left-hand) rotation matrix
having the form ($\epsilon \sim \frac{v}{\Lambda}$)
\begin{equation}
L =\left( \begin{array}{ccc}
1 & \epsilon^{2} & \epsilon^{4} \\
\epsilon^{2} & 1 & \epsilon^{4} \\
\epsilon^{4} & \epsilon^{4} & 1
\end{array} \right).
\end{equation}

The CKM matrix is equal to $U_L^{\dag}D_{L}$, where $U_L$ is the
left-hand rotation matrix for the up-quark sector and $D_L$ is the
corresponding matrix for the down-quark sector. In this case, since
all three sectors have been assumed to have identical $Z_6$ assignment
schemes, these two matrices will both be of the above form, leading to
a matrix with approximately 1 along the diagonals and a Cabbibo
angle (the first-second generation mixing angle) larger than the other
two angles, which is in qualitative agreement with observation. To
further enhance this result, one could vary some of the coupling
constants a little to split these latter two angles, however the
simplicity of taking all
three sectors to transform via the same pattern under $D$ leads to a
much more significant divergence from reality since clearly the
particles of a given generation do {\it not\ } have (approximately) the
same mass.

At this point we shall leave the simpler generation-hierarchy model we
have been using in favour of a more complicated pattern of charge
assignments with a view to explaining the intra-generational
hierarchies and to obtain better results for the mixing angles.

\section{A More Complicated Hierarchy}
\label{full-hier}

In this section we shall show how our mechanism can be used to
generate a more complicated hierarchy. The previous pattern, looked at
in Sec.~\ref{gen-gap}, had several problems with it that we would like
to fix in the more complicated hierarchy to be used here. First, it
failed to explain the hierarchy between the 1st-and-3rd generation
mixing angle and the 2nd-and-3rd generation mixing angle. Second, it
was assumed that all the masses of a given generation were of roughly
the same size. To fit with reality, one needs to assume large
splittings in the coupling constants to explain such hierarchies as
that between the top quark and the tauon
($\frac{\lambda_{\rm top}}{\lambda_{\rm tauon}} \sim 100$). Finally,
for reasonable choices for $\epsilon$ based on the observed
ratios between the masses of particles in different generations, the
Cabibbo angle generated by the simple hierarchy comes out far too
small.
The hierarchy we shall address is shown in table \ref{hiertable},
with the top quark heading the list, and ending with the electron
(similar hierarchies are looked at in \cite{bighier}).

The smallest suitable discrete group that we can use turns out to be
$Z_{13}$. Reasonable results were achieved using this
group with the quark particle fields having the charge assignments
under the discrete symmetry shown in table \ref{zed13}, providing only
that one assumes that while the coupling constants must all be of
order 1, they need not all be strictly equal to 1. Note that it
is simple to generate the right magnitudes for the masses if mixing
angles do not have to be considered, but that finding a suitable
result for the CKM matrix {\it as well as\ } for the masses can be very
difficult. Indeed, because of neutral Higgs flavour changing effects,
there must also be a suitable hierarchy in the
right-hand rotation matrices. It is in fact possible to obtain
satisfactory masses and CKM mixing angles using a $Z_{9}$ symmetry,
but in
this case the right-hand rotation matrix for the down-quark sector is
highly degenerate, which leads to extreme lower limits (of the order
of 5 TeV or more) on the Higgs masses (see Sec.~\ref{fcncs}).

The assignments given in table \ref{zed13} will generate two
4-dimensional terms in the mass lagrangian, both involving the
left-hand top quark field:
\begin{equation}
{\cal L}_{\rm 4-d} = \lambda_{tt}\overline{t}_{L}t_{R}\phi_{2} +
\lambda_{tc}\overline{t}_{L}c_{R}\phi_{2} + H.c.
\end{equation}
Leaving aside any further terms involving the above combinations of
fermion fields (which have negligible effect on the mass matrices due
to the $\epsilon^{2}$ suppression factor), the allowed dimension-6
mass terms are
\begin{eqnarray}
\Lambda^{2}{\cal L}_{\rm 6-d} & = &
\lambda_{tu}\overline{t}_{L}u_{R}\phi_{2}(\phi^{\dag}_{1}\phi_{2}) +
\lambda_{ct}\overline{c}_{L}t_{R}\phi_{2}(\phi^{\dag}_{1}\phi_{2}) +
\nonumber \\ & &
\lambda_{cc}\overline{c}_{L}c_{R}\phi_{2}(\phi^{\dag}_{1}\phi_{2}) +
\lambda_{bb}\overline{b}_{L}b_{R}\phi_{2}(\phi^{\dag}_{1}\phi_{2}) +
\nonumber \\ & &
\lambda_{\tau\tau}\overline{\tau}_{L}\tau_{R}\phi_{2}(\phi^{\dag}_{1}\phi_{2})
+
+ H.c.
\end{eqnarray}
Similarly, ignoring the (suppressed) repetitions of previous
fermion field combinations, the dimension-8 terms are
\begin{eqnarray}
\Lambda^{4}{\cal L}_{\rm 8-d} & = &
\lambda_{cu}\overline{c}_{L}u_{R}\phi_{2}(\phi^{\dag}_{1}\phi_{2})^{2} +
\lambda_{sb}\overline{s}_{L}b_{R}\phi_{2}(\phi^{\dag}_{1}\phi_{2})^{2} +
\nonumber \\ & &
\lambda_{ss}\overline{s}_{L}s_{R}\phi_{2}(\phi^{\dag}_{1}\phi_{2})^{2} +
\lambda_{\mu\mu}\overline{\mu}_{L}\mu_{R}\phi_{1}(\phi^{\dag}_{2}\phi_{1})^{2}
+
\nonumber \\ & &
\lambda_{\mu e}\overline{\mu}_{L}e_{R}\phi_{1}(\phi^{\dag}_{2}\phi_{1})^{2}
+ H.c.,
\end{eqnarray}
the dimension-10 terms are
\begin{eqnarray}
\Lambda^{6}{\cal L}_{\rm 10-d} & = &
\lambda_{uu}\overline{u}_{L}u_{R}\phi_{1}(\phi^{\dag}_{2}\phi_{1})^{3} +
\lambda_{bs}\overline{b}_{L}s_{R}\phi_{2}(\phi^{\dag}_{1}\phi_{2}) +
\nonumber \\ & &
\lambda_{sd}\overline{s}_{L}d_{R}\phi_{2}(\phi^{\dag}_{1}\phi_{2})^{2} +
\lambda_{db}\overline{d}_{L}b_{R}\phi_{1}(\phi^{\dag}_{2}\phi_{1})^{3} +
\nonumber \\ & &
\lambda_{dd}\overline{d}_{L}d_{R}\phi_{1}(\phi^{\dag}_{2}\phi_{1})^{3} + H.c.
\end{eqnarray}
the dimension-12 terms are
\begin{eqnarray}
\Lambda^{8}{\cal L}_{\rm 12-d} & = &
\lambda_{ut}\overline{u}_{L}t_{R}\phi_{1}(\phi^{\dag}_{2}\phi_{1})^{4} +
\lambda_{uc}\overline{u}_{L}c_{R}\phi_{1}(\phi^{\dag}_{2}\phi_{1})^{4} +
\nonumber \\ & &
\lambda_{bd}\overline{b}_{L}d_{R}\phi_{2}(\phi^{\dag}_{1}\phi_{2}) +
\lambda_{ds}\overline{d}_{L}s_{R}\phi_{1}(\phi^{\dag}_{2}\phi_{1})^{3} +
\nonumber \\ & &
\lambda_{e\tau}\overline{e}_{L}\tau_{R}\phi_{1}(\phi^{\dag}_{2}\phi_{1})^{4} +
\lambda_{e\mu}\overline{e}_{L}\mu_{R}\phi_{1}(\phi^{\dag}_{2}\phi_{1})^{4} +
\nonumber \\ & &
\lambda_{ee}\overline{e}_{L}e_{R}\phi_{1}(\phi^{\dag}_{2}\phi_{1})^{4}
+ H.c.,
\end{eqnarray}
with the remaining three mass-matrix entries
($\overline{\tau_{L}}e_{R}$,$\overline{\tau_{L}}\mu_{R}$ and
$\overline{\mu_{L}}\tau_{R}$) coming from dimension-14, -14 and -16
terms respectively.
%
%

In the ``up'' quark sector, the above mass terms lead to the
(undiagonalised) mass matrix

\begin{equation}
M_{U} = \left( \begin{array}{ccc}
\eta & \delta & \delta \\
\mu & m & m \\
m & M & M
\end{array} \right),
\end{equation}
where $M \sim \lambda v \gg m \sim \lambda v (\frac{v}{\Lambda})^2 \gg
\mu \sim \lambda v (\frac{v}{\Lambda})^4 \gg \eta \sim \lambda v
(\frac{v}{\Lambda})^6 \gg \delta \sim \lambda v
(\frac{v}{\Lambda})^8$. After diagonalising
this matrix, we get the mass eigenvalues for the top, charm and up
quarks shown below:
\begin{equation}
\begin{array}{ccc}
m_{t} \sim M, & m_{c} \sim m, & m_{u} \sim \eta
\end{array}
\end{equation}
and a left-hand rotation matrix of the form (again, $\epsilon \sim
\frac{v}{\Lambda}$)
\begin{equation}
L_{U} = \left( \begin{array}{ccc}
1 & \epsilon^{6} & \epsilon^{8} \\
\epsilon^{2} & 1 & \epsilon^{2} \\
\epsilon^{4} & \epsilon^{2} & 1
\end{array} \right).
\end{equation}

For the ``down'' quark sector, we get the following mass matrix (with
m, $\mu$, and so on being of the same orders of magnitude as for the
``up'' quark matrix):

\begin{equation}
M_{D} = \left( \begin{array}{ccc}
\eta & \delta & \eta \\
\eta & \mu & \mu \\
\delta & \eta & m
\end{array} \right),
\end{equation}
which leads to a left-hand rotation matrix of the form

\begin{equation}
L_{D} = \left( \begin{array}{ccc}
1 & \epsilon^{2} & \epsilon^{4} \\
\epsilon^{2} & 1 & \epsilon^{2} \\
\epsilon^{4} & \epsilon^{2} & 1
\end{array} \right).
\end{equation}
and the mass eigenstates
\begin{equation}
\begin{array}{ccc}
m_{b} \sim m, & m_{s} \sim \mu, & m_{d} \sim \eta.
\end{array}
\end{equation}

Combining these rotation matrices gives a CKM matrix with the
following form

\begin{equation}
U_{CKM} = \left( \begin{array}{ccc}
1 & \epsilon^{2} & \epsilon^{4} \\
\epsilon^{2} & 1 & \epsilon^{2} \\
\epsilon^{4} & \epsilon^{2} & 1
\end{array} \right),
\end{equation}
that is, the same as $L_{D}$. The charged leptons will have mass
eigenstates
\begin{equation}
\begin{array}{ccc}
m_{\tau} \sim m, & m_{\mu} \sim \mu, & m_{e} \sim \delta.
\end{array}
\end{equation}
The left-hand rotation matrix generated for the charged leptons
(under these particular assignmnents) has the form
\begin{equation}
L_{CL} = \left( \begin{array}{ccc}
1 & \epsilon^{4} & \epsilon^{6} \\
\epsilon^{4} & 1 & \epsilon^{8} \\
\epsilon^{6} & \epsilon^{6} & 1
\end{array} \right).
\end{equation}
These quantities will, of course, be unphysical unless the neutrinos
have mass.

The mass spectrum desired above has now been achieved, but we still
have an unwanted approximate degeneracy ($\theta_{12} \sim
\theta_{23}$) among the quark mixing angles. It appears
that this is the best one can achieve for any set of assignments
and discrete groups up to at least $Z_{13}$.

At this point, then, in order to improve on the model, one needs to
add something extra to it. One possible such addition would be
to have the VEVs of the two Higgs doublets differ (although they would
still be of the same order of magnitude, so that $v_{1} \sim v_{2} \ll
\Lambda$), so that for instance contributions to the Cabibbo angle
would come from particle fields coupling to the Higgs doublet with the
larger VEV, while the second-third generation mixing angle would
receive contributions only from fields coupled to the other Higgs
doublet. For the particular choice of fermion-field assignments under
the discrete symmetry given in this paper, however, this possibility
proved unable to resolve the angle degeneracy while maintaining
suitable values for the fermion masses.

Here, we shall look at another possibility, that of taking the
assignments under
$Z_{13}$ used above and assuming that the coupling constants are of
order 1, rather than strictly equal to 1. It turns out to be possible
to generate a reasonable mass and mixing angle hierarchy with none of
the coupling constants exceeding the range 0.3 -- 3, and with an
$\epsilon$ value of 0.24.\footnote{One may legitimately enquire as to
whether or not perturbation theory remains valid for coupling
constants that are as large as about 3. A detailed partial wave
unitarity calculation would be necessary to answer this question
rigorously. However, experience with Yukawa coupling
constants \cite{pertable}
suggests that values less than about 4 or 5 lie in the perturbative
regime.} A typical set of values
for the coupling constants is shown below. For the ``up''-quark
coupling constants:
\begin{equation}
\label{ucc}
\left( \begin{array}{ccc}
\lambda_{uu} & \lambda_{uc} & \lambda_{ut} \\
\lambda_{cu} & \lambda_{cc} & \lambda_{ct} \\
\lambda_{tu} & \lambda_{tc} & \lambda_{tt}
\end{array} \right) = \left( \begin{array}{ccc}
-1 & -0.75 & -2 \\
2.25 & 1.25 & 1.25 \\
2 & -3 & -1.25
\end{array} \right);
\end{equation}
for the ``down''-quark coupling constants:
\begin{equation}
\label{dcc}
\left( \begin{array}{ccc}
\lambda_{dd} & \lambda_{ds} & \lambda_{db} \\
\lambda_{sd} & \lambda_{ss} & \lambda_{sb} \\
\lambda_{bd} & \lambda_{bs} & \lambda_{bb}
\end{array} \right) = \left( \begin{array}{ccc}
-2 & 3 & 1 \\
-3 & 0.3 & 1.5 \\
-1 & 1 & -3
\end{array} \right);
\end{equation}
and for the charged leptons:
\begin{equation}
\label{ecc}
\left( \begin{array}{ccc}
\lambda_{ee} & \lambda_{e \mu} & \lambda_{e \tau} \\
\lambda_{\mu e} & \lambda_{\mu \mu} & \lambda_{\mu \tau} \\
\lambda_{\tau e} & \lambda_{\tau \mu} & \lambda_{\tau \tau}
\end{array} \right) = \left( \begin{array}{ccc}
2 & 0.5 & 1 \\
0.5 & 0.8 & 1 \\
-1 & 1 & 1
\end{array} \right).
\end{equation}
With these values, and taking $\epsilon = 0.24$, we find that the CKM
matrix becomes
\begin{equation}
U_{CKM} \simeq \left( \begin{array}{ccc}
0.98 & 0.20 & 0.0011 \\
0.20 & 0.98 & 0.042 \\
0.0072 & 0.041 & 1
\end{array} \right)
\end{equation}
which is clearly a good match with reality. For the masses,
taking the mass of the down quark as definite (this model generates
mass ratios, rather than absolute values for the masses), the above
values for the coupling constants generate the following spectrum:
\begin{equation}
\label{masseqn}
\begin{array}{ccc}
m_{u} \simeq 6.0\;{\rm MeV} & m_{c} \simeq 1.57\;{\rm GeV} & m_{t}
\simeq 104\;{\rm GeV} \\
m_{d} = 9.9\;{\rm MeV} & m_{s} \simeq 37.6\;{\rm MeV} & m_{b} \simeq
5.5\;{\rm GeV} \\
m_{e} \simeq 0.52\;{\rm MeV} & m_{\mu} \simeq 102\;{\rm MeV} & m_{\tau}
\simeq 1.75\;{\rm GeV.}
\end{array}
\end{equation}
These are not the only possible sets of coupling constants that
provide a spectrum like this, but there are several
qualitative constraints on the coupling constant
values. The nature of the resulting spectrum is more sensitive to some
of the coupling constants than to others. In particular, at least one
of the two dimension-4 terms involving the left-hand top quark field
must have a relatively large coupling constant [e.g. $\lambda_{tc}$ in
Eq.\ (\ref{ucc}) above], and any of the charge $-\frac{1}{3}$
coupling constants in Eq.\ (\ref{dcc}) that are not shown equal to 1
are tightly constrained to the values given. In summary, we performed
a coarse search through the $0.3 < |\lambda| < 3$ parameter space
region, and we found no promising regions other than the one in
Eqs.\ (\ref{ucc} -- \ref{ecc}) plus perturbations around it.

The mass hierarchy resulting from this set of coupling
constants is largely in good agreement with observation, with most of
the results lying within the experimentally allowed ranges
for the masses, and can thus be considered accurate to the level
at which small corrections due to higher-dimensional terms in the mass
Lagrangian could be expected to have an effect. The most obvious
mismatch is with the strange quark mass which is too low by a factor
of at least 3. One can alter various coupling
constants in such a way as to correct this problem, but only at the
expense of creating much larger discrepancies elsewhere in the
spectrum. Aside from this problem, the parameter examples given in
Eqs.\ (\ref{ucc} -- \ref{ecc}) are attractive because they use only
numbers that can be reasonably called ``of order 1'' to achieve a good
mass and mixing angle spectrum, and thus we have achieved very good
progress.

Finally, it should be noted that the above value for $\epsilon$, with
the VEV $v$ around 174 GeV, implies that the high energy scale
$\Lambda$ for this theory is around the TeV mark, not far removed
from present-day available accelerator energies.

\section{FCNCs and Other Phenomenology}
\label{fcncs}

Naturally, the introduction of a new horizontal symmetry as proposed in
this paper will lead to new channels for flavour changing neutral
currents (FCNCs). In this section, the possible FCNCs resulting from
the extra Higgs doublet will
be investigated to see what constraints observational limits on such
processes place on the Higgs masses and the high energy scale
$\Lambda$. The severest constraints come from processes such as
contributions to the $K-\overline{K}$ mass-difference and leptonic
decays. In both cases, the Higgs-induced process will be suppressed
by vertex factors of powers of $\frac{v}{\Lambda}$ due to the
suppressed nature of light weak-eigenstate Higgs interactions, and
small mixing angles if the light mass-eigenstates are first rotated
into the heavier weak-eigenstates.

There are two parts to the kaon mass difference calculation in the
standard model, the short-range and long-range contributions. Only the
short-range contribution has been successfully determined, giving an
order-of-magnitude correct result. It is assumed, therefore, that any
other contributions (including the long-range contribution and any
non-SM contributions) will be of the same order of magnitude. In
practice, this means that such contributions are limited only by the
experimental bounds on the kaon mass difference, not on the error in
this value. Nevertheless, the constraint is severe. The short-range
contribution to the kaon mass difference comes from the two
channels shown in Fig.\ \ref{kmassdiff}. Following Okun \cite{okun}, in
the standard model we have
\begin{equation}
{\cal L}_{\Delta s=2} =
G_{2}\overline{s_{L}}\gamma_{\alpha}(1+\gamma_{5})d\cdot
\overline{s_{L}}\gamma_{\alpha}(1+\gamma_{5})d
\end{equation}
where $G_{2}$ is given by
\begin{equation}
G_{2} \simeq \frac{G_{F}^{2}m_{c}^{2}}{16\pi^{2}}(\sin^{2}\theta_{c}
\cos^{2}\theta_{c})
\end{equation}
with
\begin{equation}
G_{F} = \frac{g^{2}}{2^{\frac{5}{2}}m_{W}^{2}} = 1.165 \times 10^{-5}
({\rm GeV})^{-2},
\end{equation}
where $g$ is the weak coupling-constant and $m_{W}$ is the W-boson
mass.
Measurements of the mass difference proceed via investigations of
the decay products of the kaons, and so lead to the requirement that
\begin{equation}
\label{g2eqn}
G_{2}f_{K}^{2}m_{K} \simeq 10^{-15} {\rm GeV}
\end{equation}
where $f_{K} \simeq 165$ MeV and $m_{K} = 498$
MeV ~\cite{pdg} are the kaon decay constant and mass,
respectively. In the processes shown in
Fig.\ \ref{kmassdiff}, the Higgs mass limit can be calculated from
Eq.\ (\ref{g2eqn}) by replacing $G_{2}$ with $G_{H}$, where
\begin{equation}
G_{H} = \frac{\lambda_{H}^{2}}{m_{H}^{2}}
\mbox{$\times$ Mixing-angle factors $\times 2$ (for the two diagrams).}
\end{equation}
Here, $m_{H}$ is the Higgs mass, and $\lambda_{H}$ is the coupling
strength of the Higgs particle to the fermion fields and is
proportional to the mass of the fermions involved.
The strongest coupling will be between the Higgs field and the weak
b-quark eigenstate, so the mixing angles in this case turn out to be
of order $\epsilon^{4}$ at each vertex --- factors of $\epsilon^{2}$
coming in from the mixing of each of the right- and left-handed mass
eigenstates (we note here that a model was found using a $Z_{9}$
discrete symmetry which could provide a satisfactory mass and mixing
angle spectrum, but that the right-hand mixing angles for the down
quarks were found to be highly degenerate in this model, so that the
mixing angle factor involved in this calculation was too high,
resulting in a Higgs mass lower limit of around 5 TeV).
Taking $\lambda_{H} = m_{b}/v$, where we take $m_{b}$ to be
5.5 GeV from Eq.\ (\ref{masseqn}) and $v$ is the Higgs VEV and has the
value 174 GeV, and
(from Sec.~\ref{full-hier}) $\epsilon = \frac{v}{\Lambda} = 0.24$ we
therefore get
\begin{equation}
m_{H}^{2}\geq\frac{\epsilon^{8}m_{b}^{2}f_{K}^{2}m_{K}}
{10^{-15}v^{2}}\;\;{\rm GeV}
\end{equation}
which leads to
\begin{equation}
m_{H}\geq\;\;545{\rm GeV}.
\end{equation}
This figure should be taken as an order of magnitude limit only
as the above calculation is not completely rigorous, but a lower limit
of a few hundred GeVs is nevertheless an acceptable result.

The limits resulting from the leptonic sector are much less
severe. For example, consider the lepton-number violating tau decay
\begin{equation}
\tau \rightarrow \mu\overline{\mu}e.
\end{equation}
To simplify the calculation, we take the ratio of this process to the
SM process
\begin{equation}
\tau \rightarrow \mu\overline{\nu_{\mu}}\nu_{\tau},
\end{equation}
which will eliminate the Lorentz factors associated with the
calculation. We find, noting the current limit on the lepton-number
violating decay (${\rm BR}(\tau^{-}\rightarrow\mu^{-}\mu^{-}e^{+}) <
2.7 \times 10^{-5}$ \cite{pdg}),
\begin{equation}
m_{H}^{2}\geq\frac{{\rm BR}
(\tau^{-}\rightarrow\mu^{-}\overline{\nu_{\mu}}\nu_{\tau})
m_{W}^{2}\lambda_{H}^{2}\epsilon^{6}}
{{\rm BR}(\tau^{-}\rightarrow\mu^{-}\mu^{-}e^{+})g^{2}}
\end{equation}
leading to the limit
\begin{equation}
m_{H} \geq 4.7\;\;{\rm GeV}
\end{equation}
for $g^2 = 4\pi\alpha = \frac{4\pi}{137}$, $m_{W} = 80$ GeV ~\cite{pdg},
$\epsilon = 0.24$ and $\lambda_{H} = \frac{m_{\tau}}{v} \approx 0.01$.
Since this value for the Higgs mass is much less constraining than
that from the kaon mass-difference, experimental observations on the
latter process are likely to provide the first verification (or the
strongest counter-argument) to the ideas discussed in this paper.

\section{Conclusion}
\label{conc}

We have combined effective theory with a discrete symmetry in order to
better explain the observed mass and mixing angle hierarchy. An
example of the method was given for the case of the rather simplistic
model in which the masses of the three generations are split but masses
within a given generation remain roughly equal.

The method was then applied to a more ambitious, and consequently more
realistic, hierarchy. For the assumption that all coupling constants
remain strictly equal to one, the mass hierarchy is easily explained,
but problems arise in trying to generate a realistic mixing angle
hierarchy. By weakening the restriction on the coupling constants so
that they need only be {\em of order} one, a spectrum can be produced
which matches reasonably well with observation. The results are
compatible with a high-energy scale for the theory of the order of a
TeV.

In Sec.~\ref{fcncs}, the limits placed on the Higgs
mass from FCNCs were calculated, and these place a lower limit
on the Higgs mass of a few hundred GeV, due to
constraints coming from the neutral kaon mass difference. The
constraints from leptonic FCNCs were found to be a lot less severe.

\begin{figure}
\caption{Higgs-induced contributions to the Kaon mass difference}
\label{kmassdiff}
\end{figure}

\begin{table}
\caption{Charges $e^{in\frac{\pi}{3}}$ for Fermion and Higgs Fields
Under Z-6 Symmetry}
\label{zed6}
\begin{tabular}{cc}
$n$ & Particle Fields \\ \hline
3 & $\tau_{R}$ \\
2 & $e_{R}$, $\mu_{R}$ \\
1 & $\phi_{1}$, $\mu_{L}$ \\
0 & $\phi_{2}$, $e_L$ \\
-1 & \\
-2 & $\tau_{L}$ \\
\end{tabular} \end{table}

\begin{table}
\caption{A More Complicated Approximate Hierarchy for the Fermion Masses}
\label{hiertable}
\begin{tabular}{cc}
(dim 4) & t \\
(dim 6) & b,c,$\tau$ \\
(dim 8) & s,$\mu$ \\
(dim 10) & u,d \\
(dim 12) & e \\
\end{tabular} \end{table}

\begin{table}
\caption{Charges $e^{in\frac{2\pi}{13}}$ for Fermion and Higgs Fields
Under Z-13 Symmetry}
\label{zed13}
\begin{tabular}{cc}
$n$ & Particle Fields \\ \hline
6 & $\phi_{1}$\\
5 & \\
4 & \\
3 & $b_{R}$, $u_{R}$, $(\tau,\nu_{\tau})_{L}$ \\
2 & $t_{R}$, $c_{R}$, $e_{R}$, $\mu_{R}$ \\
1 & \\
0 & $(u,d)_{L}$, $(e,\nu_{e})_{L}$ \\
-1 & \\
-2 & $s_{R}$, $\tau_{R}$, $(\mu,\nu_{\mu})_{L}$ \\
-3 & $d_{R}$ \\
-4 & \\
-5 & $(t,b)_{L}$ \\
-6 & $\phi_{2}$, $(c,s)_{L}$ \\
\end{tabular} \end{table}

\end{document}